# A Study of Language Usage Evolution in Open Source Software

Siim Karus
University of Tartu, Estonia
University of Zurich, Switzerland
siim.karus@ut.ee

Harald Gall
University of Zurich
Switzerland
gall@ifi.uzh.ch

## ABSTRACT
The use of programming languages such as Java and C in Open Source Software (OSS) has been well studied. However, many other popular languages such as XSL or XML have received minor attention. In this paper, we discuss some trends in OSS development that we observed when considering multiple programming language evolution of OSS. Based on the revision data of 22 OSS projects, we tracked the evolution of language usage and other artefacts such as documentation files, binaries and graphics files. In these systems several different languages and artefact types including C/C++, Java, XML, XSL, Makefile, Groovy, HTML, Shell scripts, CSS, Graphics files, JavaScript, JSP, Ruby, Phyton, XQuery, OpenDocument files, PHP, etc. have been used. We found that the amount of code written in different languages differs substantially. Some of our findings can be summarized as follows: (1) JavaScript and CSS files most often co-evolve with XSL; (2) Most Java developers but only every second C/C++ developer work with XML; (3) and more generally, we observed a significant increase of usage of XML and XSL during recent years and found that Java or C are hardly ever the only language used by a developer. In fact, a developer works with more than 5 different artefact types (or 4 different languages) in a project on average.

## Categories and Subject Descriptors
D.2.7 [**Software Engineering**]: Distribution, Maintenance, and Enhancement – *Restructuring, reverse engineering, and reengineering, version control*; D.3.2 [**Programming Languages**]: Language Classifications – *object-oriented languages, extensible language*; K.2 [**Computing Milieux**] History of Computing – *Software, People*

## 1. INTRODUCTION
There has been a lot of effort put into studying the use of procedural languages such as C and object-oriented languages such as Java. Even less common languages such as Perl, Python, or Ruby have received their fair share of attention. However, when looking at the statistics of most used languages, a language far more common than any of the ones mentioned earlier, strikes out. According to ohloh.net[1] which tracks more than 400,000 open source software (OSS) repositories, about 15% of actively developed OSS projects contain XML while less than 10% contain HTML, and other languages are present in less than 8 % of projects. Even more, XML is also the language with the most lines of code changed per month. The use of XML in OSS projects, however, has not received considerable attention so far.

As XML is a mark-up language, having only little meaning on its own, it would be interesting to understand, what other language it is being used with. Looking at co-evolving file types, we could investigate that issue. Even more general, the question of which languages and file types are used together and, therefore, are co-evolving in OSS projects can be formulated.

To address this research question, we studied 22 OSS software repositories over 12 years. Our study focused on two levels of file type couplings: developer and commit level. On the developer level, developers in the projects were studied regarding their language experience in the projects. For that, we addressed the following questions:

- Which languages and artefacts are commonly used in OSS development and in what proportions?
- How many file types does a developer typically work with and are there some usage patterns for file types?
- How has the language usage and, as a consequence, the language expertise requirements for developers changed during the observation period?

At the commit level, co-changing files appearing together in commits were studied. For that, we addressed the following questions:

- Which co-evolution patterns can be observed in OSS projects (e.g., are there distinct dependencies between languages or artefact types commonly edited together)?
- How have the dependencies between file types used in the projects changed during the observation period?

Additionally, on a more general level of OSS projects studied, we were interested in what are the most common languages or artefact types in the 22 OSS projects. Our observations clearly show some trends: (1) JavaScript and CSS files most often co-evolve with XSL; (2) almost every Java developer but only every second C developer works with XML; (3) over the years a significant increase of XSL and XML usage can be observed showing technological shifts due to framework development.

---
[1] http://www.ohloh.net

Table 1. Overview of OSS projects used in the study.

| Project Name | Type | Period Studied | Dev.-s | Art. Types | Rev.-s | Files |
|---|---|---|---|---|---|---|
| cocoon | business | 2003 - 2003 | 18 | 22 | 99 | 3575 |
| commons | business | 2007 - 2009 | 45 | 31 | 2981 | 6029 |
| esb | business | 2007 - 2009 | 28 | 23 | 1419 | 922 |
| httpd | business | 1996 - 1997 | 10 | 6 | 99 | 79 |
| Zope | business | 1996 - 1997 | 3 | 3 | 100 | 26 |
| wsas | business | 2007 - 2009 | 37 | 16 | 1517 | 1356 |
| wsf | business | 2007 - 2009 | 37 | 27 | 3642 | 4836 |
| bibliographic | desktop | 2003 - 2008 | 4 | 7 | 477 | 155 |
| bizdev | desktop | 2003 - 2009 | 6 | 6 | 129 | 9 |
| dia | desktop | 1997 - 2009 | 152 | 12 | 4196 | 3042 |
| docbook | desktop | 2000 - 2009 | 30 | 26 | 7540 | 6612 |
| docbook2X | desktop | 1999 - 2007 | 2 | 16 | 1082 | 304 |
| exist | desktop | 2002 - 2009 | 39 | 30 | 7116 | 4651 |
| fbug-read-only | desktop | 2007 - 2007 | 2 | 12 | 23 | 291 |
| feedparser-read-only | desktop | 2004 - 2009 | 5 | 10 | 263 | 4651 |
| gnome-doc-utils | desktop | 1999 - 2009 | 127 | 11 | 1032 | 328 |
| gnucash | desktop | 1997 - 2009 | 21 | 21 | 11757 | 3855 |
| groovy | desktop | 2003 - 2009 | 61 | 28 | 8339 | 5583 |
| nltk-read-only | desktop | 2001 - 2001 | 3 | 8 | 98 | 83 |
| subversion | desktop | 2000 - 2000 | 3 | 6 | 99 | 52 |
| tei | desktop | 2001 - 2009 | 14 | 26 | 5772 | 3832 |
| valgrind | desktop | 2002 - 2009 | 21 | 15 | 6857 | 3339 |

The paper is organized as follows. In section 2, the OSS projects used in the study are introduced and described. Section 3 details the findings about developers and Section 4 discusses our findings about co-evolution of different types of language usage in OSS projects. Threats to validity are outlined in Section 5 and related work is discussed in Section 6. We conclude with our results and give a brief outlook onto future work.

## 2. Dataset

To study development patterns, a dataset of 22 OSS projects was used. The projects were split into desktop type and business (server) type projects by their nature. That is, projects offering business functionality such as web services were considered to be business type projects and projects mainly used in desktop environments were considered to be of type desktop. Table 1 shows the periods studied, number of developers (Devs), number of different artefact types (Art.Types) used, number of commits/revisions (Revs) and files for each of the projects used in this study. The number of files stated in the table includes all files including those that were deleted during the course of the projects and are not present in the latest revision of the corresponding project.

The projects were chosen so that they would represent a wide spectrum of development projects in terms of type, duration, development team size, and usage scenario. Whilst business type projects *commons*, *esb*, *wsas*, and *wsf* belong to a larger complex super-project called WSO2 and *bizdev* and *bibliographic* are utilities for OpenOffice, the rest of the projects were mostly independent from each other. *Docbook*, *docbook2X*, and *gnome-doc-utils* represent documentation development tools. *Exist* [1], *feedparser-read-only*, *groovy*, *tei*, *subversion*, *nltk-read-only* (natural language toolkit), *fbug-read-only* (firebug) and *valgrind* are projects for software project development aids or libraries. *httpd*, *Zope*, and *cocoon* are application development platforms. *Gnucash* is an accounting application, and *dia* is a diagramming solution.

To better understand, how well the dataset represents the population, the dataset was compared with graphs publicly available from ohloh.net. In both cases the usage of C/C++ displayed steep decrease in its usage share and Java presented sudden emergence and strong yet no longer growing presence. The share of commits to XML files was increasing and had reached the highest share of file types used. The main difference between the dataset used and ohloh.net data was the lower usage of HTML in our dataset. The dataset used in the study accordingly exhibited higher share of XML and Java compared to ohloh.net data. The distribution of major artefacts worked on in the projects in out dataset during different years is shown on Figure 1.

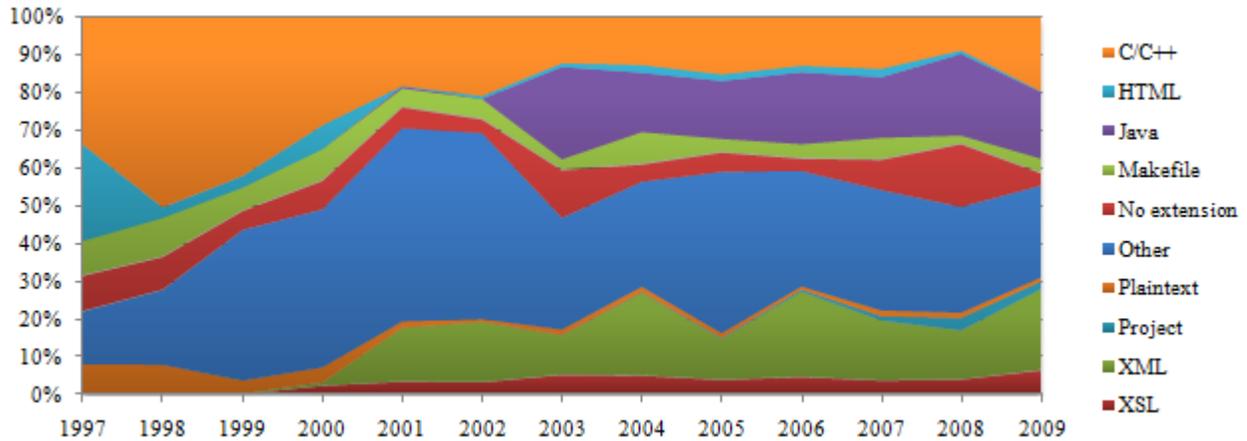

Figure 1. Distribution of major file types worked on in projects per year.

We identified and classified 45 major file types of the most common file extensions in our repository: Archive, Audio, awk, Binary, C, C#, C++, Command Script, CSS, Data, DTD, Graphics, Groovy, HTML, Java, JavaScript, jsp, Makefile, Manifest, MS Office, No extension, OpenDocument, OpenXML, Patch and Diff, PDF, Perl, PHP, Plaintext, PostScript, Project, Properties, Python, Resources, Rich Text, Ruby, sed, Shell Script, SQL, SQML, TeX, WSDL, XML, XML Schema, XQuery, and XSL. Other languages were present with only very few files.

Most of the files were classified by their extensions; however, there were some exceptions:

- The category plaintext includes files with the extensions .txt, .readme, .changes, .install and files named "README", "INSTALL", "TODO", "COPYING", "COPYRIGHT", "AUTHORS", "LICENSE", "ACKNOWLEDGEMENTS", "NEWS", "NOTES", "ChangeLog", and "CHANGES". These files contain project documentation in plain text format.

- The category Project contains XML files with root element "project". These files are mostly used by IDEs to store project configuration or by the build tools (e.g. ant or Maven) to store project build configuration. The distribution of these subtypes is shown in Figure 3.

- The category Manifest contains files with the extension ".manifest" and files named "manifest.xml".

- The category Properties contains files with extension "properties" and XML files with root element "properties". These files are used in Java projects to store application configuration.

- The category "Perl" additionally contains extensionless text files which begin with "#!/usr/bin/perl".

- The category "Shell Scripts" additionally contains extensionless text files which begin with "#!/bin/sh", "#!/bin/sh", or "#!/bin/bash".

- The category "SGML" additionally includes "catalog" files.

Every file can belong only to one category at once. For example, XSL, XAML, XHTML, etc. files were not counted as XML files, neither are files that are included in other categories due to exceptions (e.g. files named "manifest.xml", which belong to category Manifest). Another special general group is "files without extensions", which includes folders due to the differences in how the repositories present their data.

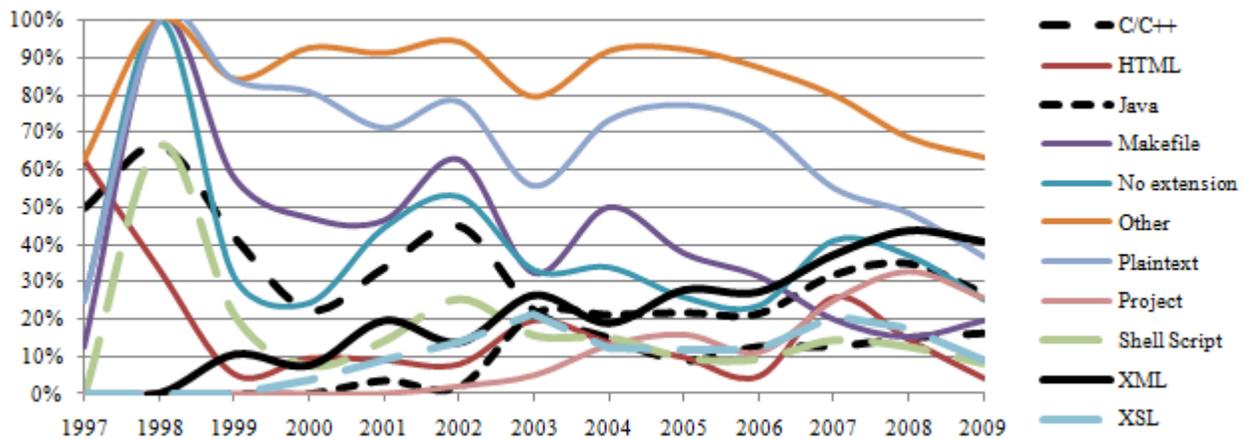

Figure 2. Proportion of developers generating different types of artefacts during different years.

TABLE 2.        MOST ABUNDANT FILE TYPES USED TOGETHER BY DEVELOPERS.

|  | Archive | C/C++ | CSS | Graphics | Groovy | HTML | Java | JavaScript | Makefile | No extension | Perl | PHP | Plaintext | Project | Properties | Shell Script | WSDL | XML | XML Schema | XSL |
|---|---|---|---|---|---|---|---|---|---|---|---|---|---|---|---|---|---|---|---|---|
| Archive |  | 4% | 32% | 42% | 32% | 55% | 95% | 34% | 9% | 91% | 11% | 11% | 70% | 80% | 73% | 43% | 27% | 92% | 32% | 42% |
| C/C++ | 3% |  | 8% | 25% | 2% | 38% | 13% | 2% | 82% | 79% | 25% | 21% | 85% | 21% | 7% | 58% | 16% | 52% | 9% | 18% |
| CSS | 27% | 10% |  | 67% | 4% | 80% | 62% | 60% | 29% | 84% | 14% | 11% | 76% | 66% | 49% | 50% | 34% | 90% | 27% | 74% |
| Graphics | 27% | 25% | 53% |  | 4% | 71% | 56% | 47% | 35% | 88% | 16% | 11% | 75% | 60% | 43% | 50% | 36% | 84% | 26% | 64% |
| Groovy | 48% | 4% | 8% | 8% |  | 24% | 94% | 4% | 0% | 72% | 4% | 0% | 50% | 76% | 52% | 20% | 0% | 48% | 2% | 4% |
| HTML | 24% | 24% | 41% | 47% | 7% |  | 52% | 35% | 32% | 82% | 20% | 14% | 67% | 55% | 37% | 47% | 34% | 72% | 22% | 45% |
| Java | 34% | 7% | 27% | 31% | 23% | 43% |  | 30% | 7% | 71% | 5% | 6% | 52% | 72% | 50% | 30% | 21% | 79% | 17% | 38% |
| JavaScript | 34% | 3% | 73% | 73% | 3% | 82% | 85% |  | 15% | 84% | 11% | 8% | 66% | 85% | 55% | 42% | 45% | 89% | 35% | 82% |
| Makefile | 3% | 35% | 10% | 15% | 0% | 21% | 6% | 4% |  | 50% | 16% | 9% | 92% | 9% | 4% | 28% | 6% | 33% | 4% | 17% |
| No extension | 22% | 29% | 24% | 32% | 12% | 46% | 47% | 20% | 42% |  | 15% | 11% | 76% | 44% | 32% | 41% | 18% | 65% | 13% | 32% |
| Perl | 16% | 57% | 27% | 37% | 4% | 71% | 22% | 16% | 88% | 96% |  | 27% | 90% | 27% | 16% | 71% | 8% | 65% | 14% | 37% |
| PHP | 24% | 73% | 30% | 39% | 0% | 76% | 39% | 18% | 73% | 100% | 39% |  | 91% | 67% | 18% | 73% | 45% | 79% | 27% | 52% |
| Plaintext | 11% | 20% | 15% | 18% | 5% | 25% | 23% | 11% | 52% | 51% | 9% | 6% |  | 24% | 17% | 26% | 10% | 35% | 8% | 19% |
| Project | 34% | 14% | 34% | 39% | 22% | 55% | 86% | 36% | 13% | 79% | 7% | 13% | 63% |  | 52% | 40% | 30% | 82% | 22% | 42% |
| Properties | 49% | 7% | 40% | 44% | 23% | 58% | 93% | 37% | 9% | 89% | 7% | 5% | 70% | 81% |  | 50% | 32% | 90% | 31% | 50% |
| Shell Script | 24% | 48% | 33% | 42% | 7% | 60% | 47% | 23% | 56% | 93% | 26% | 18% | 90% | 52% | 41% |  | 30% | 76% | 23% | 41% |
| WSDL | 33% | 30% | 52% | 68% | 0% | 98% | 73% | 55% | 28% | 93% | 7% | 25% | 80% | 88% | 60% | 68% |  | 95% | 50% | 67% |
| XML | 25% | 22% | 30% | 36% | 9% | 47% | 61% | 25% | 32% | 75% | 12% | 10% | 62% | 53% | 37% | 38% | 21% |  | 15% | 41% |
| XML Schema | 59% | 24% | 59% | 73% | 2% | 93% | 85% | 63% | 27% | 100% | 17% | 22% | 95% | 93% | 83% | 76% | 73% | 98% |  | 73% |
| XSL | 24% | 16% | 52% | 57% | 2% | 60% | 60% | 47% | 35% | 77% | 14% | 13% | 67% | 57% | 43% | 43% | 31% | 85% | 23% |  |

The data was gathered in May 2009 and contains revision information from February 1996 to April 2009 (12 years).

## 3. Developers

To study the habits of developers and find language usage sets commonly present in the projects, we extracted developer information from revision data in the revision control systems (CVS and SVN). We then listed the file types used by each developer and analyzed the data.

### 3.1 Languages Used

The most popular artefact type used by 64% of developers was identified as plaintext files. Files without extensions (mostly changes to directory structure) were edited by 37% of developer. Makefiles and XML were used by 34% of the developers, making these artefact types share the third and the fourth position. Java files were edited by 26% of developers, followed by the popularity of project files (21%) and HTML files (19%). Surprisingly, C/C++ files were used by fewer developers (14%) than XSL files (15%). Considering that XSL has gained popularity while C/C++ has lost its, it can be said that in the more recent years there are more active XSL developers than there have been C/C++ developers. The ratio of developers using different file types throughout the study period is shown in Figure 2. Note that year 2009 figures only account for the first quarter of the year (data collection point).

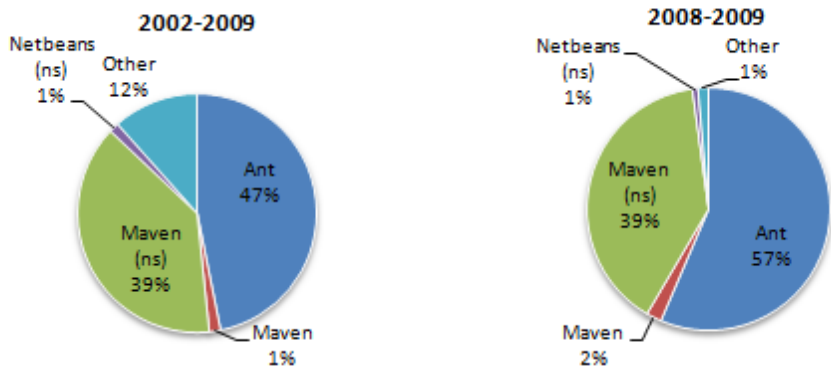

Figure 3. Distribution of project file subtypes in the dataset and the last years (ns marks revisions with explicit namespace).

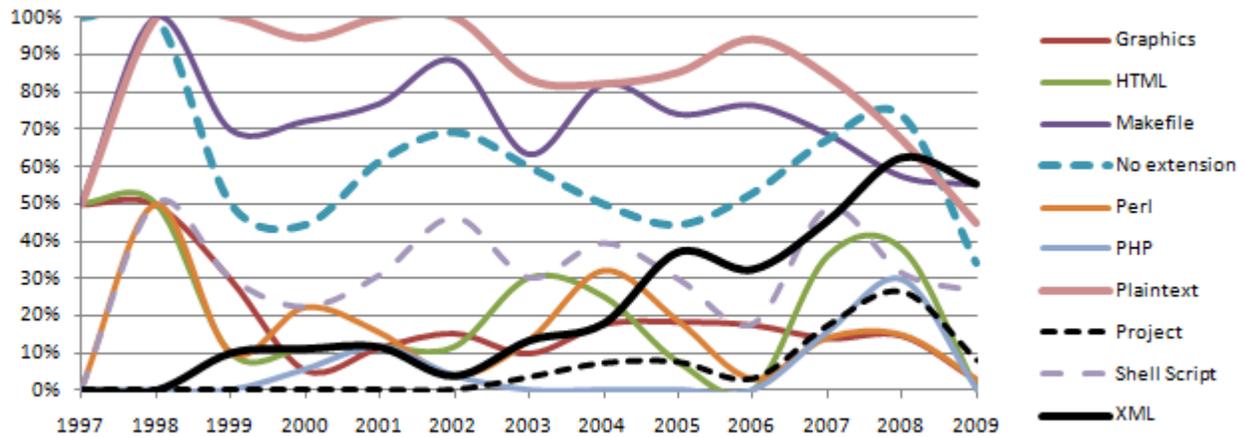

Figure 4. Additional languages used in at least 20% of commits by C developers.

Table 2 displays the most common artefacts commonly used by the same developer. The most common combination of file types used by a developer was Java and XML. This can be explained by both of these languages being in the top four artefact types encountered.

Among language pairs used by more than 10% of developers, Plaintext and XML were the most popular second languages. Interestingly, only 85% of XSL developers modified XML files. This could be caused by XSL applied on either XML files with extensions other than "xml" or by XSL being used to transform documents created by the runtime or received from third party. XML Schema editors were also active in XML development in 98% of cases (followed by WSDL, Archive file, Properties file, JavaScript and CSS developers). XSL was commonly seen together with languages used in web development (i.e. HTML, XML Schema, CSS, JavaScript and graphics files).

One should keep in mind that making commits to certain type of files does not necessarily mean that the developer has expertise in the responding field. The commits could be deferred from other developers or be just necessities solved with the help of other developers. The identification of expertise is a complex task studied in other works like [2] and [3].

These sets of commonly co-appearing languages along with the popularity of languages allow us to identify major classes of developers by the languages they use. The three major classes defined by the most popular languages are C/C++ developers, Java developers and XML developers.

### 3.1.1 C/C++ Developers.

C/C++ developers were frequent users of plaintext files (used by 85% of C/C++ developers) and Makefiles (82%). This is expected as plaintext files were commonly used to document C/C++ projects while Makefiles were the chosen technology to control the C/C++ build process. Files without extensions were modified by 79% of C/C++ developers, which can be explained by a decent folder structure. The fourth most common language used by C developers was Shell scripts (58% of C developers) followed closely by XML (52% of C developers).. Further details about most abundant file types used by developers can be seen in Table 2. The matrix shows for developers using file type specified in rows the percentage of developers also using file type specified by column (e.g. 42% of developers of archive files also worked with XSL files).

During the late 1990s, most developers had worked with C, written Makefiles and created some other types of artefacts. Since then, less than half of the developers have written C or C++ code (after dropping to 9% in 2005, the percentage of developers using C or C++ has climbed steadily to 16% in 2009) and Makefiles have been continuously become less popular dropping from 63% in 2002 to 19% in 2009 (see Figure 2).

The most commonly used language by C developers (apart from C itself) has almost always been Makefiles and most commonly used file type "plaintext" (see Figure 4). Nevertheless, the popularity of plaintext files is slowly decreasing among C developers. Use of XML has made a strong impression since its adoption in 1998 and has reached more than 60% of C developers in 2005. This could be related to more widespread adoption of XML standards and XML replacing Makefile based building environments.

### 3.1.2 Java Developers.

A total of 79% of Java developers also worked on XML files, making XML the most popular language used together with Java. The second most popular language used together with Java was Project files, which was used by 72% of Java developers. The top three also includes files without extensions *(directory structure modifications)*, which were used by 71% of Java developers. The next popular file types were used significantly less (see Table 2).

Usage of Java has been on the rise with more than 30% of developers having used it in the last study period. As shown in Figure 5, Java developers use more different types of artefacts than C/C++ developers. The graph also displays that Java developers are writing XSL by themselves less frequently than they used to. As XSL has become more popular in general, it can be explained by XSL being written by developers more focused on XSL and less on Java

It is wrong to assume that the popularity of XML in Java projects is mainly due to project build files. In fact, more than half of the .xml files found in Java projects were of project-specific types. Also, the use of binary files (including .class and .jar files) by Java developers has dropped below 20%. This could be a result of using separate library repositories instead of having all files in revision control repository.

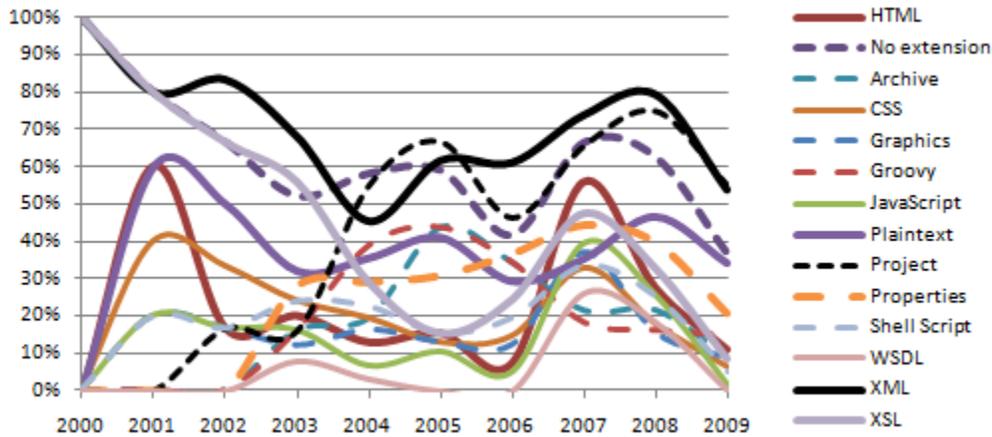

Figure 5. Additional languages used in at least 20% of commits by Java developers.

### 3.1.3 *XML* Developers.

Knowledge of developing XML files has also been on a steady rise with more than 40% of the developers having used it in the period 2008-2009.

XML developers come from different areas and work with variety of different artefacts. This is shown by the fact that only there are lots of different artefact types used in more than 20% of commits by XML developers (Figure 6). The most popular file type modified by XML developers is "files without extensions" (used by more than 60% of XML developers), which has been slowly losing its popularity among XML developers since 2002.

It is a good practice for .xml files to have explicitly defined namespace(s), which can be used to verify the files. However, 31% of .xml files did not specify their namespace. The most often encountered namespaces were http://www.tei-c.org/ns/1.0 (29%) and http://docbook.org/ns/docbook (17%), showing that XML is often used for project or domain specific languages. This is confirmed by the most popular root elements: <refentry> (either in http://docbook.org/ns/docbook namespace or no namespace specified, 27% of all root elements) and <elementSpec> (in http://www.tei-c.org/ns/1.0 namespace, 21% of all root elements).

The most common root element in XML files used with Java was <project> (8% of XML files). These files were classified into "Project" category which mostly contained files without explicit namespace (see Figure 3).

XSL, which is commonly accompanied by XML, has been used by a steady 10% of developers since its introduction in 2000 with another 10% gain since 2007.

## 3.2 The First Commit

While commits in general tell us about the file types used in the projects, the first commit made by a developer tells a lot about the initial experience and start-out of developers. We can expect developers to prefer file types and languages they are more familiar with when joining the development team.

The first commit also shows the patterns of how developers get involved or build up their contribution. It is expected that a developer using more different languages in its first commit needs to understand the project's architecture and build practices better than a developer who starts by just changing a few lines in a single file.

The number of different types of files in the developers' first commit was usually less than four with almost half being commits of a single file type. There is also a trend of using fewer files and file types in the first commit towards the end of the study period (see Figure 7).

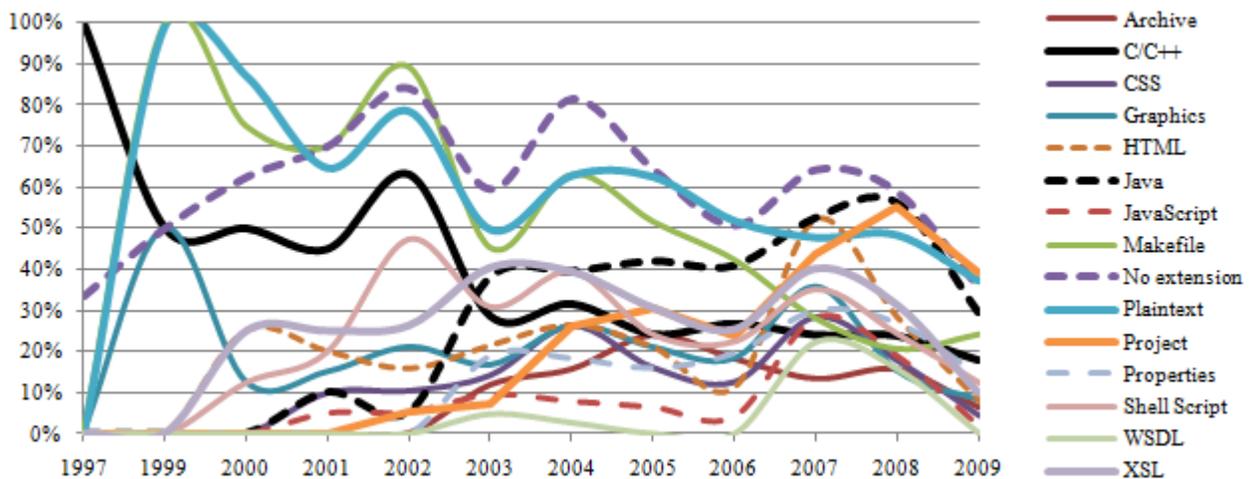

Figure 6. Additional languages used in at least 20% of commits by XML developers.

In the 1990s C and Makefile were the most popular choices as the first commit language, but in the later years XML and Java have taken the lead. The number of different file types present in the first commit and the number of files in developers' first commit are lower in 2008-2009 than in the 1990s. In 2008-2009 75% of first commits were of files of single type whilst only 15% of commits made in 1990-s were of single file type.

The analysis of the most common file type combinations shows that 12% of first commits included both Makefiles and files without extensions (3% contained no other file types). Most common single file type commits were for files without extension (8%), Java (8%), and XML (7%).

When considering all commits, 21% contained both Makefiles and files without extensions (only 2% contained only these two and 2% contained also C files). Java and XML files were encountered in 4% of all commits (1% contained only these). The most common single type commit file types were Java, XML, HTML and XSL (all accounted for only 1% of commits). This shows that developers expand their competences by learning and deploying new languages during the project; however, these languages will be tightly coupled causing files of different file types to be changed at a time.

Most developers started with XML (16%), Makefiles (17%) or Java (15%), but rarely used only one language in their first commit.

## 4. Co-Changes

To find out, which file types or languages have been used together and which files are co-changed, the file types for each commit were analyzed. The common file types committed together were identified for both project types business and desktop separately.

### 4.1 Business Type Projects

The most commonly encountered combination present in business projects was a combination of Java and XML. About 15% of all commits made to Java files were accompanied by changes to XML files and 32% of changes to .xml files co-occurred with changes to Java files.

The strongest bidirectional relation was found to be between JavaScript and XSL files. More than 40% files of these types were co-changed. Also, in 42% of cases, a change to a CSS file was accompanied by a change to an XSL file. Other co-occurrences were much less frequent even in case of web file types (e.g. changes to CSS files were accompanied with changes to Javascript files in only 36% of the cases). One reason for that could be that in the projects studied XSL was mainly used for generating reports and data presentations in web applications. This usually results in XSL being used in place of writing HTML directly and thus gets changed often during user interface development and testing. On the other hand, co-change rate of less than 45% with other presentation type artefacts (Javascript, CSS or graphics files) indicates that business type projects in the dataset used XML for business document transformations about as often as they used these for generating presentations.

Commits, which contained multiple files of the same type were most frequently commits of files of type graphics (53% of all graphics commits), C (52% of all C commits), XML Schema (50%), PHP (48%), Java (47%) and binary and XSL (both 43%). That is, 53% of commits of graphics files contained more than one graphics file. XML Schema files were more often changed along with Java or XML files than any other XML Schema file. Similarly WSDL files were changed with XML or Java files more often than with any other WSDL file. Web file types such as CSS and JavaScript were more often committed with XSL files than with any other of the same kind. This means that graphics developers, C/C++, and XML Schema developers are more likely to work in patches than other developers.

We also found that changes to binary files (e.g. .class, .o and .dll files) were on average accompanied by changes to files in almost four other file types while commits to Java files were accompanied by the average of 0.57 artefacts of other types. This could be caused by developers committing compiled files along with source code. Frequently co-changing file types also include XML Schema (3.7 other file types), and WSDL (2.4 file types). Files usually not co-changing were of types PHP (0.1 other file types), C (0.4 other file types), and Java (0.6 other file types).

The most commonly encountered file types in multiple file type commits are in order of frequency: Java (on average, present in 27% of commits with files of other type), Project (15% of commits with files of other type), XML (13%), and XSL (11%), and files without extensions (11%). The details of co-changes in business type projects can be seen in Table 3. The table shows, how many commits containing artefacts of the type listed in row header contained artefacts of the type listed in columns.

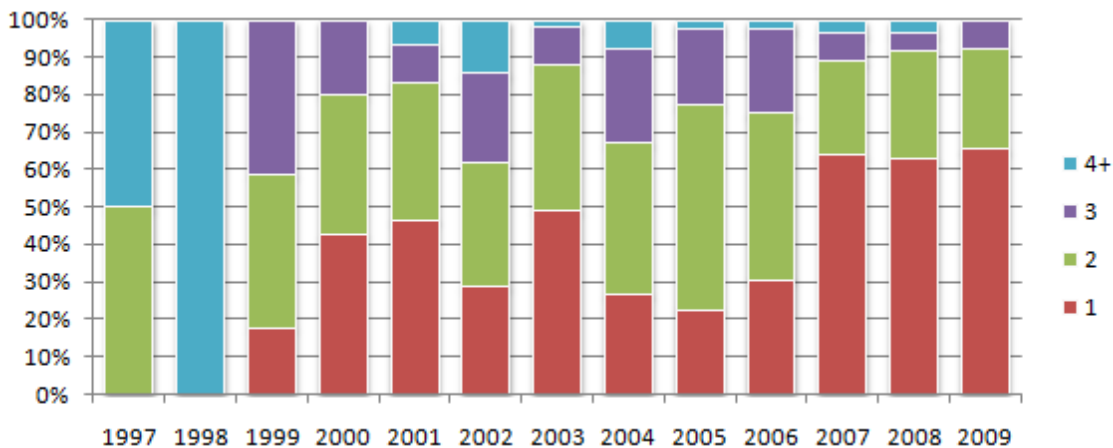

Figure 7. Number of file types in developer first commit by year.

TABLE 3. ARTEFACT TYPES COMMIT TOGETHER IN BUSINESS TYPE PROJECTS.

| | C/C++ | CSS | Graphics | HTML | Java | JavaScript | Makefile | No extension | OpenDocument | PHP | Plaintext | Project | Properties | Ruby | Shell Script | WSDL | XML | XSL |
|---|---|---|---|---|---|---|---|---|---|---|---|---|---|---|---|---|---|---|
| C/C++ | 52% | 0% | 1% | 1% | 0% | 0% | 9% | 5% | 0% | 4% | 2% | 0% | 0% | 2% | 4% | 1% | 3% | 0% |
| CSS | 0% | 2% | 14% | 19% | 8% | 36% | 0% | 5% | 0% | 0% | 1% | 7% | 2% | 0% | 1% | 1% | 7% | 42% |
| Graphics | 2% | 26% | 53% | 22% | 9% | 7% | 2% | 12% | 0% | 0% | 3% | 11% | 5% | 0% | 3% | 2% | 11% | 14% |
| HTML | 0% | 9% | 6% | 21% | 10% | 19% | 1% | 7% | 0% | 0% | 3% | 7% | 1% | 0% | 1% | 1% | 10% | 16% |
| Java | 0% | 1% | 1% | 3% | 47% | 6% | 0% | 9% | 0% | 0% | 1% | 14% | 4% | 0% | 1% | 2% | 15% | 5% |
| JavaScript | 0% | 12% | 1% | 14% | 15% | 28% | 0% | 2% | 0% | 0% | 0% | 6% | 1% | 0% | 0% | 1% | 6% | 45% |
| Makefile | 10% | 0% | 1% | 1% | 0% | 0% | 18% | 5% | 0% | 0% | 2% | 1% | 1% | 0% | 10% | 1% | 3% | 0% |
| No extension | 2% | 2% | 2% | 5% | 22% | 2% | 2% | 31% | 0% | 1% | 3% | 17% | 4% | 3% | 5% | 3% | 19% | 2% |
| OpenDocument | 0% | 0% | 0% | 1% | 1% | 0% | 0% | 4% | 40% | 0% | 0% | 0% | 0% | 0% | 0% | 0% | 0% | 0% |
| PHP | 4% | 0% | 0% | 0% | 0% | 0% | 0% | 2% | 0% | 48% | 0% | 0% | 0% | 0% | 2% | 2% | 2% | 0% |
| Plaintext | 7% | 2% | 4% | 12% | 11% | 2% | 5% | 23% | 0% | 1% | 19% | 17% | 6% | 1% | 8% | 4% | 27% | 4% |
| Project | 0% | 2% | 1% | 3% | 26% | 4% | 0% | 12% | 0% | 0% | 2% | 20% | 3% | 0% | 2% | 3% | 30% | 4% |
| Properties | 1% | 3% | 5% | 6% | 65% | 5% | 1% | 27% | 0% | 0% | 5% | 23% | 20% | 0% | 7% | 4% | 33% | 3% |
| Ruby | 6% | 0% | 0% | 1% | 1% | 2% | 1% | 20% | 0% | 0% | 1% | 1% | 0% | 46% | 2% | 2% | 3% | 2% |
| Shell Script | 4% | 0% | 1% | 2% | 8% | 1% | 11% | 12% | 0% | 0% | 3% | 8% | 3% | 1% | 15% | 2% | 11% | 0% |
| WSDL | 4% | 3% | 4% | 9% | 43% | 8% | 4% | 34% | 0% | 9% | 8% | 41% | 8% | 2% | 8% | 26% | 47% | 13% |
| XML | 1% | 2% | 2% | 6% | 32% | 5% | 1% | 16% | 0% | 1% | 3% | 33% | 5% | 0% | 4% | 3% | 27% | 6% |
| XSL | 0% | 14% | 2% | 11% | 12% | 44% | 0% | 2% | 0% | 0% | 1% | 5% | 1% | 0% | 0% | 1% | 7% | 43% |

**Co-change trends**. In summary, we have observed the following language usage trends in our dataset: (1) Java and XML files co-evolve most often compared to the other file and language types, whereas C files rarely co-evolve with any other file type; (2) Binary files co-evolve with XML and Java files in most of the cases; (3) WSDL files often co-change with Java and XML files; (4) JavaScript files co-evolve with XSL files; (5) XML Schema files co-changes with WSDL, XML and Java files; and (6) XSL files basically only co-change with JavaScript files.

## 4.2 Desktop Type Projects

In desktop projects, C was historically the most common language in our dataset. As such, files representing languages and file types related to C development were commonly changed together. For example, changes to C files were accompanied by changes to files without extensions (e.g. folders, Linux executables) in 32% of the cases and with changes to Makefiles in 14% of the cases. A similar observation was made with graphics files, which were committed together with Makefiles in 29% of the cases and with changes to files having no extensions in 40% of the cases. As opposed to business type projects, changes to Java files were accompanied by changes to XML files only in 3% of the cases. Details of these co-occurrences can be seen in Table 4. The most common co-change pattern was observed for Groovy files, which were co-changed with Java files in about half of the cases.

Graphics commits had the most diversity of co-changed artefacts (2.6 other file types were committed with graphic files on average),
followed by command scripts (2.2 file types), JavaScript (1.9 file types), and Binary files (1.9 file types). Binary files were committed together with Java or XML files in more than third of the cases. The most independent file types were Java (co-changed with 0.4 file types on average) and XSL (co-changed with 0.4 file types on average).

Multiple file type commits most often contained files without extensions, XML, Java, XSL and C files. Multiple graphics files were co-committed in 62% of the cases (i.e. 62% of commits with graphics files contained more than one graphics file). Other file types often changed in a bulk (i.e. with multiple files in a commit) were C (48% of all C commits), Binary (47%), PHP (46%), and Java (42%) files.

**Co-change trends.** In summary, we have observed the following language co-change trends: (1) Binary files co-change with Java and XML files; (2) C files with Makefiles; (3) Command Scripts with Makefiles, Shell Script and XML files; (4) CSS files with XML; (5) Groovy with Java files; (6) JavaScript with CSS and XSL files; and (7) Ruby files co-change with XSL.

One of the major differences to business (server) type projects is that in desktop OSS projects we observed much lower co-evolution of Java and XML files (in either direction the co-change was half as likely as in business type projects). On the other hand, CSS files co-changed with XML files twice as often in desktop projects. Other trends are similar for both OSS project types investigated.

## 5. Threats to validity

Threats to the validity of our work are confounding and selection (bias and generalisability).

TABLE 4. ARTEFACT TYPES COMMIT TOGETHER IN DESKTOP TYPE PROJECTS.

| | Archive | C/C++ | CSS | DTD | Graphics | Groovy | HTML | Java | Makefile | No extension | Perl | PHP | Plaintext | Project | Properties | Python | Shell Script | XML | XQuery | XSL |
|---|---|---|---|---|---|---|---|---|---|---|---|---|---|---|---|---|---|---|---|---|
| Archive | 41% | 0% | 3% | 1% | 2% | 11% | 3% | 34% | 0% | 14% | 2% | 0% | 6% | 23% | 17% | 1% | 9% | 22% | 2% | 8% |
| C/C++ | 0% | 48% | 0% | 0% | 0% | 0% | 1% | 0% | 14% | 32% | 0% | 0% | 1% | 0% | 0% | 0% | 0% | 1% | 0% | 0% |
| CSS | 2% | 0% | 20% | 1% | 4% | 0% | 4% | 4% | 10% | 15% | 1% | 3% | 2% | 4% | 2% | 1% | 2% | 19% | 11% | 36% |
| DTD | 1% | 4% | 1% | 21% | 2% | 0% | 1% | 2% | 6% | 23% | 2% | 0% | 3% | 1% | 1% | 0% | 1% | 41% | 0% | 7% |
| Graphics | 3% | 18% | 8% | 3% | 60% | 1% | 9% | 4% | 29% | 40% | 4% | 0% | 11% | 3% | 3% | 2% | 11% | 16% | 2% | 9% |
| Groovy | 2% | 0% | 0% | 0% | 0% | 29% | 1% | 51% | 0% | 3% | 0% | 0% | 4% | 7% | 3% | 0% | 0% | 1% | 0% | 0% |
| HTML | 1% | 7% | 2% | 0% | 2% | 2% | 21% | 9% | 6% | 10% | 1% | 1% | 3% | 4% | 1% | 3% | 1% | 8% | 0% | 2% |
| Java | 1% | 0% | 0% | 0% | 0% | 13% | 1% | 43% | 0% | 2% | 0% | 0% | 2% | 4% | 2% | 0% | 1% | 3% | 1% | 1% |
| Makefile | 0% | 29% | 1% | 0% | 1% | 0% | 1% | 0% | 25% | 41% | 1% | 0% | 2% | 0% | 0% | 1% | 2% | 7% | 0% | 5% |
| No extension | 0% | 34% | 1% | 1% | 1% | 1% | 1% | 1% | 21% | 14% | 1% | 0% | 2% | 1% | 1% | 2% | 1% | 4% | 0% | 5% |
| Perl | 2% | 12% | 2% | 2% | 4% | 0% | 3% | 3% | 20% | 21% | 14% | 0% | 6% | 3% | 2% | 1% | 4% | 10% | 1% | 11% |
| PHP | 0% | 3% | 6% | 0% | 0% | 0% | 4% | 0% | 6% | 23% | 0% | 44% | 0% | 0% | 0% | 0% | 6% | 1% | 0% | 17% |
| Plaintext | 2% | 16% | 1% | 1% | 2% | 8% | 3% | 16% | 11% | 18% | 2% | 0% | 17% | 8% | 3% | 1% | 4% | 6% | 1% | 3% |
| Project | 5% | 0% | 2% | 0% | 1% | 12% | 3% | 25% | 0% | 9% | 1% | 0% | 7% | 17% | 12% | 1% | 2% | 9% | 2% | 4% |
| Properties | 13% | 0% | 3% | 0% | 2% | 17% | 3% | 34% | 1% | 15% | 1% | 0% | 8% | 39% | 14% | 1% | 4% | 21% | 3% | 6% |
| Python | 1% | 10% | 2% | 0% | 1% | 0% | 8% | 1% | 9% | 46% | 1% | 0% | 3% | 2% | 1% | 16% | 2% | 14% | 0% | 2% |
| Shell Script | 6% | 10% | 2% | 0% | 5% | 0% | 2% | 10% | 19% | 32% | 3% | 3% | 8% | 6% | 4% | 1% | 26% | 11% | 2% | 7% |
| XML | 1% | 3% | 2% | 3% | 1% | 1% | 2% | 6% | 7% | 9% | 1% | 0% | 1% | 2% | 2% | 1% | 1% | 29% | 1% | 10% |
| XQuery | 2% | 0% | 17% | 0% | 2% | 0% | 1% | 19% | 1% | 6% | 1% | 0% | 4% | 8% | 4% | 1% | 3% | 19% | 28% | 12% |
| XSL | 0% | 0% | 4% | 0% | 0% | 0% | 0% | 1% | 6% | 10% | 1% | 1% | 1% | 1% | 0% | 0% | 1% | 10% | 1% | 30% |

Confounding is an internal threat to the explanations given to some observations. That is, there might be some event in the society that have changed the characteristics of developers or languages used (e.g. companies campaigns to push their technologies), that we can not directly relate to the dataset, which makes these relationships difficult or impossible to identify. The impact of these events might end up attributed to some other change we could find correlation with. This threat cannot be avoided.

Selection threat is both internal (bias) and external (generalisability). It is internal as the selection might be biased towards certain projects (e.g. by motivation). We do accept that the dataset studied has somewhat elite collection of projects as there were no single developer projects, which account to about a half of the population of all OSS projects [4]. We have validated the representativeness of our dataset against the data provided by ohloh.net and found the general characteristics of these datasets to be similar despite the threat of bias. We found no differences in the artefact popularity rankings and the biggest difference observed was the popularity of HTML code. This similarity gives high confidence to the generalisability and representativeness of the results of this study.

## 6. Related Work

The idea of studying cross-file co-changes has been addressed by some research so far. However, these studies have been often language specific and they rarely look at different file types. Even studies encompassing multiple file types have been limited to specific file types. For example, Zimmerman et al. studied how lines of different files evolve in a project [5]. Their study is limited to textual files and focused more on visualisation and clustering of files based on their change history.

Weißgerber et al. have built a plug-in for Eclipse to show how likely different files are to be changed together [6]. Their tool does not exclude any files. However, they aim to visualise patterns emerging in specific projects regarding the co-evolution of files. They do not try to describe co-evolution on file type or artefact type level. As such their tool is useful for monitoring software development processes. In contrast, our paper explains more general patterns spanning through OSS software projects.

Dattero et al. conducted a survey during 2000-2001 and looked into differences by the developer gender [7]. They discovered that female developers are more likely to work with deprecated technologies. They also found that female developers tend to be less experienced and are familiar with only 2.53 languages as opposed to 3.25 languages male developers were familiar with. These numbers are similar to our findings, however, we also saw that the average number of different file types (usually representing different technologies) used by developers has decreased during the period studied.

The different patterns of evolution of OSS have been outlined by Nakakoji et al. [8]. They determined that there are three main types of OSS: exploration-oriented, utility-oriented, and service-oriented. These types determine how the software evolves and how the developers behave. The projects studied here spanned over all these types – business projects being largely exploration-oriented, *gnucash*, *bibliographic* being utility-oriented and *eXist*, *feedparser-read-only* service-oriented (as in providing stable services not to be confused with SOA). The study also shows that projects have a development speed cycle, along which the projects transform from one type to another. This can be used to explain the fluctuations in the language and file type shares over time as seen in this study.

Open-source software repositories have been used for studying various aspects of software development like developer role identification (core or associate) [9], framework hotspot detection [10] and other. These works are complementary and help developing a better understanding of sotware development process and open-source software. It has also been shown that the number and size of open-source projects are growing exponentially and open-source projects are becoming more diverse by expanding into new domains [11].

## 7. Conclusions and Future Work

We investigated the revision data of 22 OSS projects and tracked the evolution of multiple programming language usage. Our findings can be summarized from a language and a developer perspective.

First, as for multiple programming language usage, our study confirmed the ohloh.net data that the most popular (i.e. widely used) language in OSS software projects is XML followed by Java and C. XML has increased its popularity steadily over the last decade while C has lost its high share to various other languages of which Java has been among the more popular ones. Despite becoming popular in just a few years, Java has not been able to grow its share significantly during the last years. XSL has maintained its share for the last years.

The most commonly co-evolving files are usually of the same type. These are, ranked in order of co-evolution intensity: Java and XML; C and plaintext files; and C and Makefiles. The most co-dependent pair of file types in the business type projects studied was JavaScript and XSL with a co-change rate (measured in common commits) of more than 40% of the cases. Java and XML files (especially those of project specific types) are more likely to be edited by the same person than Java files and project definition files.

Based on the projects analyzed, we found that XSL is important for both generating user interfaces and for document transformations.

Second, as for developers, we found that fewer file types are used by new developers in their first commits, even though most developers began with experience with multiple file types. Most developers worked with at least five different file types during the period studied. 80% of Java developers worked with XML files while only 40% of C developers did so (60% in the later years).

The study of languages used by developers from 1997 to 2009 showed the decreasing importance of Makefiles and plaintext files for C developers while the importance of XML increased with almost any other language. Whilst document type definition language was being deprecated, XML Schema did not seem to replace it (neither did any other language), implying that standardised schemas are being preferred over project specific ones.

From the characteristics of developer language usage, we saw that not just knowing multiple languages is required from the developers, but developers must also understand different coding paradigms (e.g. procedural and object-oriented languages are often used side-by-side with rule and template based extensible languages). While in the 1990s they needed to know how to code in C and write Makefiles, the increased variety of languages used in newer projects and lack of distinct leaders in languages introduced the need to be familiar with multi-language development.

Future work will address to better describe the population by including newer/future projects. The ideal dataset would have more similar characteristics to the data available from ohloh.net (e.g. more HTML code), which has currently the biggest analysed listing of open source projects and as such is closest to representing the population. It is not feasible to incorporate all projects listed by ohloh.net as the data that would need to be analysed would exceed our capabilities of processing it in timely fashion.

## 8. Acknowledgments

This research was conducted during a visit of the first author to the software evolution and architecture lab at the University of Zurich. We thank the members of the lab for their valuable advice. The work is also partially funded by ERDF via the Estonian Centre of Excellence in Computer Science.